\documentclass[12pt]{article}
\begin{document}
\title{Second Quantized F-P Ghost States}
\author{E. Kazes
~\\~\\~\\
Department of Physics\\
104 Davey Laboratory\\
University Park, PA 16802}
\maketitle
\begin{abstract}
Negative norm Hilbert space state vectors can be BRST invariant, we show in a simplified Y-M model, which has gluons and ghost fields only, that such states can be created by starting with gluons only. 
\end{abstract}

\newpage
\addtolength{\baselineskip}{\baselineskip}

The covariant quantization of vector potentials is known to require a Hilbert space with indefinite metric.  In electrodynamics, states of negative metric are separated from the physical sector, which contains transverse photons through the Gupta-Bleuler construction.  In non-abelian gauge theories, path integral techniques supplemented with Faddeev-Popov\cite{fp} ghosts, and further extended with BRST\cite{brs,tiv} techniques, provide powerful computational tools.  Kugo\cite{ko,ko1}, Ojima\cite{ko,ko1,no} and Nakanishi\cite{no} investigated extensively the Yang-Mills theory and the BRST structure of its Hilbert space.  In particular they pointed out the importance of having a semi-definite Hilbert space realization of the physical sector of this theory.  In that case, states of zero norm are orthogonal to all states of semi-definite norm.  In this way zero norm states which proliferate in this model, as well as in quantum electrodynamics, are conveniently disposed of.  However, the coherence of this scheme requires that physical states, which have positive norm, not evolve dynamically into states of negative norm.  We examine the perturbative evolution of some physical states of positive norm into states of negative norm, using the second quantized F-P model, thereby confirming that BRST invariance alone is not enough for a theory to be physically admissible\cite{fs}.  We have not examined the same problem in other models such as\cite{ko,ko1,no}.  The transition amplitude to states of zero norm is also given for comparison.

The minimal Y-M Lagrangian for gluons, ghosts and auxiliary fields is
\begin{equation}
{\cal L} = {\cal L}_{o}(A)+{\cal L}_{GF}+{\cal L}_{FP}\\
\end{equation}
\begin{equation}
{\cal L}_{o}=-\frac{1}{4}F^{a}_{\mu\nu}F^{\mu \nu a}\\
\end{equation}
\begin{equation}
{\cal L}_{GF}=-\frac{1}{2}(\partial_{\mu}A^{\mu a})^2\\
\end{equation}
\begin{equation}
{\cal L}_{FP} = - i \partial^{\mu} \overline{C}^{a} (D_{\mu}C)^a\\
\end{equation}
\begin{equation}
D_{\mu} = \partial_{\mu} - i g A_{\mu}\\
\end{equation}

The lowest order transition amplitude from an initial state with two gluons to a final ghost, anti-ghost state requires two types of Feynman diagrams. The one, with two gluon-ghost vertices gives the amplitude
\begin{equation}- \frac{i}{2}g^{2} \left[p_{1} \cdot \varepsilon (k_{1}) p_{2} \cdot \varepsilon(k_{2})  C_{a_{1}b_{1}f}  C_{a_{2}b_{2}f} + p_{1} \cdot \varepsilon (k_{2}) p_{2} \cdot \varepsilon(k_{1})   C_{a_{2}b_{1}f}  C_{a_{1}b_{2}f} \right]\\
\end{equation}

The other, which has one gluon-ghost and a three gluon vertex, gives
\begin{equation}\frac{ig^{2}}{2k_{1}\cdot k_{2}} \left[\varepsilon(k_{1}) \cdot \varepsilon (k_{2}) (k_{1}-k_{2}) \cdot p_{1} - 2 p_{1} \cdot \varepsilon (k_{1}) k_{1} \cdot \varepsilon(k_{2}) + 2 p_{1} \cdot \varepsilon (k_{2}) k_{2} \cdot \varepsilon (k_{1}) \right] C_{a_{1}a_{2}f} C_{b_{1}b_{2}f}\\
\end{equation}

where $C_{abc}$ are the structure constants of the group $SU(n)$. Repeated indices are summed.  The incident gluon momenta, polarization and color indices respectively are\\
\noindent
$k_{1},\varepsilon(k_{1}),a_{1}; k_{2}, \varepsilon(k_{2}),a_{2}$.  The ghost, antighost momenta, and color indices are $p_{1}, b_{1};p_{2},b_{2}$.  A common feature of 
ghost production amplitudes is their gauge dependence, this will be seen later to follow directly from BRST invariance.  The replacement $\varepsilon(k)\rightarrow\varepsilon(k)+\lambda k$ changes the sum of (6) and (7), whereas ghost free amplitudes are invariant under this residual gauge transformation.

From the anticommutation relations of ghost and antighost fields it follows that the final state in this process
\begin{equation}
\Psi=\overline{c}_{b_{1}}^{+}(p_{1})c_{b_{2}}^{+}(p_{2})\mid  0 \rangle\\
\end{equation}
has zero norm if $p_{1}\neq p_{2}$.  Whereas for smeared ghost, antighost states
\begin{equation}
\Psi^{'}=\int \underline{d}p_{1} \underline{d}p_{2} f(p_{1}) g(p_{2}) \overline{c}_{b_{1}}^{+}(p_{1})c_{b_{2}}^{+}(p_{2})\mid 0 \rangle\\
\end{equation} 
the norm is
\begin{equation}
\langle \Psi^{'} \mid \Psi^{'} \rangle = - \mid \int f(p_{1}) g(p_{2}) \underline{d}p_{1} \underline{d}p_{2} \mid^{2} \delta_{b_{1}b_{2}}\\
\end{equation}
A negative norm results when $b_{1}=b_{2}$ for an appreciable overlap of the two momentum distribution $f$ and $g$.  Therefore the sum of (6) and (7) is the amplitudes of a zero norm vector in Fock space.

The amplitude in Eq.(7) is ambiguous for $p_{1} \rightarrow p_{2}$ since then $k_{1} \rightarrow k_{2} \rightarrow p_{2}$.  We avoid this by including a gluon in the final state with momenta, polarization and color index $k_{3}, \varepsilon (k_{3}), a_{3}$ respectively.  To obtain all amplitudes of order $g^{3}$ requires the quartic and cubic gluon-gluon vertices as well as a gluon-ghost interaction.  For $b_{1}=b_{2}=b$ the amplitude which involves the quartic vertex vanishes.  For $p_{1}=p_{2}=p$ the diagram that has three gluon-ghost vertices gives
\begin{displaymath}
\frac{g^{3}}{4} \left[ \frac{1}{p \cdot k_{1} \, \, p \cdot k_{3}} p \cdot \varepsilon (k_{1}) p \cdot \varepsilon (k_{3})  (k_{1} + k_{3}) \cdot \varepsilon (k_{2})  C_{a_{1}bf} C_{a_{2}fg} C_{a_{3}bg}
\right .
\end{displaymath}
\begin{displaymath}
+ \frac{1}{p \cdot k_{2} \, \, p \cdot k_{3}} p \cdot \varepsilon (k_{2}) p \cdot \varepsilon (k_{3})  (k_{2} + k_{3}) \cdot \varepsilon (k_{1})  C_{a_{2}bf} C_{a_{1}fg} C_{a_{3}bg}
\end{displaymath}
\begin{equation}
\left .
- \frac{1}{p \cdot k_{1} \, \, p \cdot k_{2}} p \cdot \varepsilon (k_{1}) p \cdot \varepsilon (k_{2})  (k_{1} - k_{2}) \cdot \varepsilon (k_{3})  C_{a_{1}bf} C_{a_{3}fg} C_{a_{2}bg} \right]
\end{equation}

where all repeated indices except $b$ are summed.  In order to show that the product of the three structure constant in Eq.(11) does not vanish identically it is convenient to sum over $b$, and we obtain $C_{a_{1}a_{2}a_{3}}$.
\newpage
For the same process, the amplitude that contains a gluon-ghost and the cubic gluon interaction gives
\begin{displaymath}
\frac{g^{3}}{2} \, \frac{p \cdot \varepsilon(k_{1})}{p \cdot k_{1} \,\, k_{2} \cdot k_{3}} \left[ \varepsilon(k_{2}) \cdot \varepsilon(k_{3}) \, (k_{2} + k_{3}) \cdot p - 2 p \cdot \varepsilon(k_{3}) k_{3} \cdot \varepsilon (k_{2})
\right .
\end{displaymath}
\begin{displaymath}
\left .
- 2p \cdot \varepsilon(k_{2}) k_{2} \cdot \varepsilon(k_{3}) \right] C_{a_{1}br} C_{rbl} C_{la_{2}a_{3}}
\end{displaymath}
\begin{displaymath}
+ \frac{g^{3}}{2} \, \frac{p \cdot \varepsilon(k_{2})}{p \cdot k_{2} \,\, k_{1} \cdot k_{3}} \left[ \varepsilon(k_{1}) \cdot \varepsilon(k_{3}) \, (k_{1} + k_{3}) \cdot p - 2 p \cdot \varepsilon(k_{3}) k_{3} \cdot \varepsilon (k_{1})
\right .
\end{displaymath}
\begin{displaymath}
\left .
- 2p \cdot \varepsilon(k_{1}) k_{1} \cdot \varepsilon(k_{3}) \right] C_{a_{2}br} C_{rbl} C_{la_{1}a_{3}}
\end{displaymath}
\begin{displaymath}
+ \frac{g^{3}}{2} \, \frac{p \cdot \varepsilon(k_{3})}{p \cdot k_{3} \,\, k_{2} \cdot k_{1}} \left[ \varepsilon(k_{2}) \cdot \varepsilon(k_{1}) \, (k_{2} - k_{1}) \cdot p + 2 p \cdot \varepsilon(k_{1}) k_{1} \cdot \varepsilon (k_{2})
\right .
\end{displaymath}
\begin{equation}
\left .
- 2p \cdot \varepsilon(k_{2}) k_{2} \cdot \varepsilon(k_{1}) \right] C_{a_{3}br} C_{rbl} C_{la_{2}a_{1}},
\end{equation}
and since
\begin{equation}
\sum_{b,r} \, C_{mbr} \, C_{mbr} \, \, \alpha \, \delta_{mn}
\end{equation}
it follows that the products of the structure constants is not zero for every $a_{1}, a_{2}, a_{3}$ and $b$.  Eqs.(12) and (13) give the transition amplitude to a ghost, antighost state of negative norm.  Just as in the previous example the answer is gauge dependent.

In covariant gauges, the gauge independence of transition amplitudes between physical states is known to follow from BRST symmetry, in contrast to this the transition amplitude from physical states to ghost states depends on the gauge.  The gauge dependence in Eqs.(6), (7), (11) and (12) is a non-perturbative result arising the BRST invariance of physical wave functions.  For instance, for an infinitesimal transformation $\delta$,
\begin{equation}
\delta \langle o \mid T \left(\overline{C}^{a_{1}}(x_{1}) \, A^{\mu_{2},a_{2}}(x_{2}) \, \partial_{\mu_{3}}A^{\mu_{3},a_{3}}(x_{3}) \, \partial_{\mu_{4}}A^{\mu_{4},a_{4}}(x_{4}) \right) \mid o \rangle=o
\end{equation}
from which it follows that
\begin{displaymath}
\langle o \mid T \left( \delta \overline{C}^{a_{1}}(x_{1}) \, A^{\mu_{2},a_{2}}(x_{2}) \, \partial_{\mu_{3}}A^{\mu_{3},a_{3}}(x_{3}) \, \partial_{\mu_{4}}A^{\mu_{4},a_{4}}(x_{4}) \right) \mid o \rangle
\end{displaymath}
\begin{equation}
=\langle o \mid T \left(\overline{C}^{a_{1}}(x_{1}) \, \delta A^{\mu_{2},a_{2}}(x_{2}) \, \partial_{\mu_{3}}A^{\mu_{3},a_{3}}(x_{3}) \, \partial_{\mu_{4}}A^{\mu_{4},a_{4}}(x_{4})\right) \mid o \rangle
\end{equation}
where
\begin{displaymath}
\delta \overline{C}^{a} = - {i} \, \partial^{\mu} \, A^{a}_{\mu}
\end{displaymath}
\begin{equation}
\delta A^{a}_{\mu} = \left(D_{\mu} \, C \right)^{a}
\end{equation}
are the standard BRST variations. Letting $x^{o}_{1} , x^{o}_{2} \rightarrow +\infty$ and $x^{o}_{3}, x^{o}_{4} \rightarrow - \infty$,
we use the LSZ scattering formalism.  The lhs of Eq.(15) is proportional to the gluon-gluon scattering amplitude.  The rhs is proportional to the gluon-gluon $\rightarrow$ ghost-antighost amplitude.  The l.h.s. is not zero since only three gluons are contracted with their momenta.  This shows again that the gauge dependence of what was found in Eqs.(6) and (7) is not an artifact of the weak coupling approximation.

In the first of the two cases we dealt with, the evolution from a physical state to a gauge dependent zero norm state is not a problem if that state is orthogonal to all physical states of positive norm.  However, the production of a ghost-antighost state of negative norm by starting with a two gluon state of positive norm is a serious difficulty for this  operator formulation of QCD.  Nevertheless, the ghost and anti-ghost which were introduced into the path integral quantization of gauge theories as fictitious particles  restores locality and the calculational convenience of Feynman diagrams.


\newpage
\begin{thebibliography}{99}
\bibitem{fp}Faddeev, L.D. and Popov, V.N., Phys. Lett. 25B, 29 (1967).
\bibitem{brs}Bechi, C., Rouet, A. and Stora, R., Ann. Phys. 98, 287 (1976).
\bibitem{tiv}Tyutin, I.V., Lebedev preprint FIAN No 39 (1975).
\bibitem{ko}Kugo, T. and Ojima, I., Phys. Lett. 73B, 459 (1978).
\bibitem{ko1}Kugo, T. and Ojima, I., Prog. Theor. Phys. Suppl. No. 66, I (1979).
\bibitem{no}Nakanishi, N. and Ojima, I. Covariant Operator Formalism of Gauge Theories and Quantum Gravity.  (World Scientific Singapore, 1990).
\bibitem{fs}Faddeev, L.D. and Slavnov, A.A., Gauge Fields.  (Addison-Wesley Publishing Co. 1988). See pg. 170.
\end{thebibliography}
\end{document}